\documentclass{ws-procs9x6}

\newcommand{\Ref}[1]{Ref.~\citen{#1}}
\newcommand{\Refs}[1]{Refs.~\citen{#1}}
\newcommand{\andRef}[1]{and~\citen{#1}}
\newcommand{\Fig}[1]{Fig.~\ref{#1}}
\newcommand{\Sec}[1]{Sec.~\ref{#1}}
\newcommand{\Tab}[1]{Table~\ref{#1}}
\newcommand{\Eq}[1]{Eq.~(\ref{#1})}

\newcommand{\eg}{{\it e.g.}}
\newcommand{\ie}{{\it i.e.}}
\newcommand{\etal}{{\it et al.}}

\newcommand{\BR}[1]{\ensuremath{{\rm BR}(#1)}}
\newcommand{\bra}[1]{\ensuremath{\left<#1\right|}}
\newcommand{\ket}[1]{\ensuremath{\left|#1\right>}}
\newcommand{\mean}[1]{\ensuremath{\left<#1\right>}}
\newcommand{\order}[1]{\ensuremath{\mathcal{O}(#1)}}
\newcommand{\SN}[2]{\ensuremath{#1\times10^{#2}}}

\newcommand{\stat}{\ensuremath{_\mathrm{st}}}
\newcommand{\syst}{\ensuremath{_\mathrm{sy}}}
\newcommand{\theo}{\ensuremath{_\mathrm{th}}}

\renewcommand{\dot}{\ensuremath{\mathbf{\cdot}}}
\newcommand{\Vusf}{\ensuremath{|V_{us}|f_+(0)}}

\begin{document}

\title{KAON PHYSICS: RECENT EXPERIMENTAL PROGRESS}

\author{MATTHEW MOULSON}

\address{Laboratori Nazionali di Frascati, 00044 Frascati RM, Italy\\
E-mail: moulson@lnf.infn.it}

\begin{abstract}
Numerous recent measurements of kaon decays are described, 
with an emphasis on results offering constraints on 
chiral perturbation theory calculations. An up-to-date
estimate of \Vusf\ based on semileptonic kaon 
decay rates is presented.    
\end{abstract}


\bodymatter

\section{Introduction}

The last three years have been marked by very rapid progress in 
experimental studies of kaon decays. This review is an attempt to
summarize those results that offer direct comparison for
the predictions of chiral perturbation theory (ChPT), including in
particular
the determination of the $I=0,2$ $\pi\pi$ scattering lengths from
$K\to3\pi$ and $K_{e4}$ data,
recent results on a selection of rare and radiative kaon decays, and
the determination of \Vusf\ from $K_{\ell3}$ rates.

\section{Kaon Decays and the $\pi\pi$ Scattering Lengths}

\subsection{$K^\pm\to\pi^\pm\pi^0\pi^0$ decays}

\begin{figure}
\begin{center}
\psfig{file=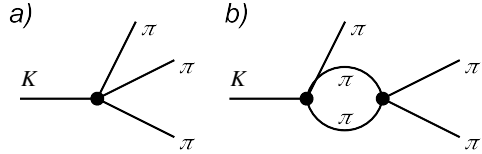,width=2.5in}
\end{center}
\caption{$K\to3\pi$ amplitudes used in the description of the cusp at
$M_{\pi^0\pi^0}^2 = 4m_{\pi^+}^2$ in the treatment of \Ref{Cab04:cusp}.}
\label{fig:cusp}
\end{figure}

In the $M_{\pi^0\pi^0}^2$ distribution for \SN{23}{6} 
$K^\pm\to\pi^\pm\pi^0\pi^0$ decays from 2003 data,
NA48/2 observes a cusp at
$M_{\pi^0\pi^0}^2 = 4m_{\pi^+}^2$ \cite{NA48+06:cusp}. 
Cabibbo \cite{Cab04:cusp} has explained the cusp in terms of the interference 
between two amplitudes: the direct amplitude illustrated in \Fig{fig:cusp}a,
and the rescattering amplitude illustrated in \Fig{fig:cusp}b.
The latter 
amplitude is proportional to $a_0-a_2$, the difference between the $I=0,2$ 
$\pi\pi$ scattering lengths; this quantity can thus be determined from
fits to the $M_{\pi^0\pi^0}^2$ distribution.
In the NA48/2 analysis, the fits are based on an extension of Cabibbo's 
original treatment, which includes contributions from the relevant 
two-loop diagrams \cite{CI05:cusp}. 
The theoretical 
contribution to the uncertainty on the determination of $a_0-a_2$ is estimated 
to be 5\%. The NA48/2 fits account for a possible contribution from pionium 
formation in the vicinity of the cusp, but the seven bins nearest the position 
of the cusp are excluded from the fit to avoid possible bias arising from
the lack of radiative corrections in the model used. NA48/2 obtains a 
fit to the $M_{\pi^0\pi^0}^2$ distribution with $\chi^2/{\rm ndf} = 146/139$
(32.5\%)
giving $(a_0 - a_2)m_{\pi^+} = 0.268(10)\stat(4)\syst(13)\theo$.
This result compares favorably to the prediction of Colangelo 
\etal\ \cite{CGL00:pipi}, based on a matching procedure between
representations of the $\pi\pi$ scattering amplitude from  
\order{p^6} ChPT calculations and from the Roy equations as
constrained by $\pi\pi$ scattering data at higher energies:
$(a_0 - a_2)m_{\pi^+} = 0.265(4)$. NA48/2 statistics will increase by a 
factor of five when all 2003--2004 data
are analyzed, which should allow the experimental uncertainty on $a_0-a_2$  
to be reduced to a level comparable to the uncertainty on the predicted value.
A precise comparison will require a reduction of the
uncertainty arising from the theoretical description of the cusp. 
Recent work on a nonrelativistic effective field theory treatment by
Colangelo \etal\cite{C+06:cusp,Kub06:Chiral} 
is promising in this regard. 
This scheme provides consistent power counting and allows electromagnetic
corrections to be included in a standard way. 
NA48/2 is currently
working on fits to the cusp using this treatment.\cite{GL06:Chiral} 

\subsection{$K_{e4}$ decays}

$K_{e4}$ decays ($K\to\pi\pi e\nu$) also provide an opportunity to study
the $\pi\pi$ interaction down to threshold. 
Fits to the kinematic 
distributions for $K_{e4}$ decays provide sensitivity
to the axial form factors $F$ and $G$, and the vector form factor $H$. 
Recent work\cite{E865+03:Ke4,BD06:QCD}
makes use of the parameterization of \Ref{AB99:Ke4FF}.
In this parameterization, the form factors are expanded in partial
waves, \eg, 
$F = F_se^{i\delta_0^0} + F_pe^{i\delta_1^1}\cos\theta_\pi$,
where $\theta_\pi$ is the angle between the $\pi^+$ momentum 
in the $\pi\pi$ system and the momentum of the $\pi\pi$ system 
in the kaon rest frame, and
$\delta_{L=0}^{I=0}$ and $\delta_{L=1}^{I=1}$ are the
$\pi\pi$ phase shifts. The coefficients $F_s$, $F_p$, etc., are expanded
in powers of $q^2=(M_{\pi\pi}^2 - 4m_\pi^2)/m_\pi^2$, so that, \eg,
$F_s = f_s + f_s'q^2 + f_s''q^4 + f_e(M_{e\nu}^2/4m_\pi^2) + \cdots$.
Fits to the kinematic distributions thus allow determination of
the parameters $f_s$, $f_s'$, etc., which can be used to 
constrain ChPT couplings ($F$ and $G$ have been calculated to two 
loops in ChPT\cite{ABT00:Ke4FF}), as well as values for 
$\delta_0^0 - \delta_1^1$ in bins of $M_{\pi\pi}$.

The Roy equations can be used to relate the $\pi\pi$ phase shifts 
to $a_0$ and $a_2$ (see, \eg, \Ref{A+01:pipi}),
but measurements of $\delta_0^0 - \delta_1^1$ provide very 
little constraint on the value of $a_2$. However, consistency with
the Roy equations and $\pi\pi$ scattering data above 0.8~GeV restricts
possible values of $a_0$ and $a_2$ to a region in the $a_0$-$a_2$ plane
often called the universal band (UB). The UB constraint can be used to 
eliminate $a_2$ in fits to $K_{e4}$ data.

The most recent published data on $K^\pm\to\pi^+\pi^-e\nu$ decays are from 
BNL E865.\cite{E865+03:Ke4}. Using a sample 
of 388k $K^+\to\pi^+\pi^-e^+\nu$ decays, E865 measures 
${\rm BR} = \SN{4.11(11)}{-5}$.
For the kinematic analysis,
the data are divided into 28800 bins (six in $M_{\pi\pi}^2$). 
For each bin in $M_{\pi\pi}^2$, the expansion parameters for
the form factors $F$, $G$, and $H$ are obtained, as well as 
values for $\delta_0^0 - \delta_1^1$. A separate fit for $a_0$ is performed
without binning the data in $M_{\pi\pi}^2$.
From this latter fit, with the application of the UB constraint, E865
obtains $a_0m_{\pi^+} = 0.228(12)\stat(4)\syst(^{+12}_{-16})\theo$, 
which compares well with the prediction from \Ref{CGL00:pipi},
$a_0m_{\pi^+} = 0.220(5)$.

\begin{figure}
\begin{center}
\psfig{file=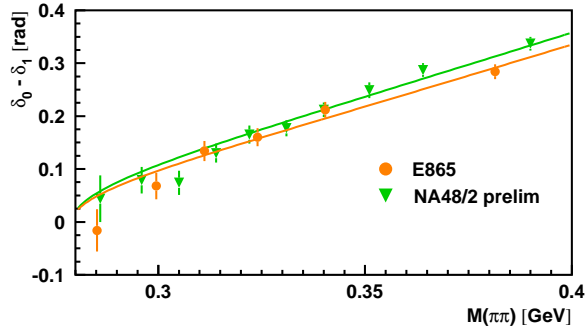,width=3in}
\end{center}
\caption{Measurements of $\delta_0^0 - \delta_1^1$ from 
E865\cite{E865+03:Ke4} and NA48/2\cite{BD06:QCD,GL06:Chiral}
(NA48/2 data are preliminary). The solid curves illustrate the Roy
equation solutions as parameterized in \Ref{A+01:pipi}, with the values
for $a_0$ quoted by each experiment. Adapted from a plot in 
\Ref{BD06:QCD}.}
\label{fig:Ke4}
\end{figure}

NA48/2 has recently obtained preliminary results on  $K^\pm\to\pi^+\pi^-e\nu$
decays as well.\cite{BD06:QCD,GL06:Chiral}.
The NA48/2 $K^\pm_{e4}$ sample includes 235k $K^+$ and 135k $K^-$
events. The data are divided into 15000 bins (ten in $M_{\pi\pi}$).
Like E865, NA48/2 measures the expansion parameters
for the form factors $F$, $G$, and $H$, as well as $\delta_0^0 - \delta_1^1$,
in bins of $M_{\pi\pi}$. With the UB constraint, NA48/2 obtains
$a_0 = 0.256(8)\stat(7)\syst(18)\theo$, which is in marginal 
($\sim$$1.7\sigma$) agreement with the prediction from \Ref{CGL00:pipi}.
The data from the two experiments are compared in \Fig{fig:Ke4}.

For either experiment, the experimental and theoretical contributions to 
the error on $a_0$ are 4--5\% and 5--7\%, respectively. 
In addition to 
the UB constraint, E865 uses the tighter constraint on $a_0$ vs.\ $a_2$ 
from \Refs{CGL01:pipi}
\andRef{CGL01:Ke4} to obtain a value with a greatly reduced theoretical 
uncertainty:  $a_0m_{\pi^+} = 0.216(13)\stat(4)\syst(2)\theo$. 
NA48/2 also expects to obtain values with smaller theoretical uncertainties
using alternate treatments.

NA48/2 has also obtained preliminary results in the channel 
$K^\pm\to\pi^0\pi^0e^\pm\nu$.\cite{BD06:QCD}
In this channel, 
Bose symmetry considerations imply that only the $L=0$ partial wave 
contributes.
With 37k events from 2003 and 2004 data, NA48/2 has obtained values for the 
form-factor parameters $f_s'/f_s$ and $f_s''/f_s$ that are 
consistent with the results obtained for $K^\pm\to\pi^+\pi^-e^\pm\nu$.

\section{Rare and Radiative Kaon Decays}

\subsection{Radiative $K_{\ell3}$ decays}

The amplitudes for radiative kaon decays can be divided into two 
components: an internal bremsstrahlung (IB) amplitude arising from 
radiation from external charged particles, and a direct-emission (DE),
or structure-dependent (SD), amplitude arising from radiation 
from intermediate hadronic states. The components
are isolated by analyzing the energy spectrum of the 
radiative photon; the study of the SD component can provide information 
about intermediate hadronic states in the decay.

For radiative $K_{\ell3}$ decays, a convenient observable is 
\begin{displaymath}
R_{\ell3\gamma}(E_{\rm min},\theta_{\rm min}) =
\frac{\BR{K_{\ell3\gamma}, E_\gamma > E_{\rm min}, \theta_{\ell\gamma} >
\theta_{\rm min}}}{\BR{K_{\ell3(\gamma)}}},
\end{displaymath}
where $E_{\rm min}$ and $\theta_{\rm min}$ are cuts on the energy
of the radiated photon and on its angle of emission (in the kaon CM frame)
with respect to the momentum of the lepton, and $\BR{K_{\ell3(\gamma)}}$
signifies the inclusive branching ratio.
These cuts are dictated by experimental
necessity or by convention, bearing in mind that the IB amplitudes 
diverge for $E_\gamma\to0$.

For $K_L\to\pi\mu\nu\gamma$, there is a recent measurement
from KTeV\cite{KTeV+05:Kl3g}: 
$R^0_{\mu3\gamma}(E_\gamma>30~{\rm MeV}) = 0.209(9)\%$. This is 
in good agreement with the earlier result from NA48\cite{NA48+98:Km3g}: 
$R^0_{\mu3\gamma} = 0.208(26)\%$, as well
as the predictions of \Ref{FFS70:Kl3g}.  
For $K_L\to\pi e \nu\gamma$, the situation is more complicated. 
Precise ChPT predictions are available---Gasser \etal\cite{G+05:Ke3g} 
have performed an \order{p^6}
ChPT calculation to obtain $R^0_{e3\gamma}(E_\gamma>30~{\rm MeV},
\theta_{e\gamma}>20^\circ) = 0.96(1)\%$.
On the experimental side, in 2001 KTeV published \cite{KTeV+01:Kl3g}
the result
$R^0_{e3\gamma} = 0.908(8)(^{+13}_{-12})\%$; the data have recently been
reanalyzed using more restrictive cuts that provide better control over
systematic effects, but which reduce the statistics by a factor of three.
The new KTeV result \cite{KTeV+05:Kl3g} is
$R^0_{e3\gamma} = 0.916(17)\%$. 
Both results are at variance with the value obtained in
\Ref{G+05:Ke3g}.
On the other hand, NA48\cite{NA48+05:Ke3g} obtains 
$R^0_{e3\gamma} = 0.964(8)(^{+11}_{-9})\%$, which is in good agreement
with \Ref{G+05:Ke3g}. KLOE has announced a preliminary 
value\cite{Mar06:Chiral} based on 20\% of its total data set:
$R^0_{e3\gamma} = 0.92(2)(2)\%$.
As the errors on this result are reduced, KLOE 
will be able to comment on the apparent discrepancy between 
the KTeV and NA48 results.

For $K^\pm\to\pi^0\mu^\pm\nu\gamma$ decays, experimental results 
have become available only recently. KEK E470 recently 
published \cite{K470+06:Km3g} the measurement 
$\BR{K^\pm_{\mu3\gamma}, E_\gamma>30~{\rm MeV}, 
\theta_{\mu\gamma}>20^\circ} = \SN{2.4(5)(6)}{-5}$, obtained 
using stopped $K^+$s. The ISTRA+ experiment at Protvino has
used in-flight $K^-$ decays to obtain measurements 
\cite{ISTRA+05:Km3g,Duk06:ICHEP}
of $\BR{K^\pm_{\mu3\gamma}}$
for $5 < E_\gamma < 30~{\rm MeV}$ and $ 30 < E_\gamma < 60~{\rm MeV}$.
All of these results are in good agreement with \order{p^4}
ChPT estimates \cite{BLC02:Kl3g,BEG93:Kl3g}.
For $K^\pm\to\pi^0 e^\pm\nu\gamma$ decays, ISTRA+ has the 
preliminary result \cite{Duk06:ICHEP}
$\BR{K^\pm_{e3\gamma}, E_\gamma>30~{\rm MeV}, 
\theta_{e\gamma}>20^\circ} = \SN{3.05(2)}{-4}$,
which is also in agreement with ChPT estimates.

\subsection{Radiative $K\to\pi\pi$ decays}

For the decay $K^\pm\to\pi^\pm\pi^0\gamma$, the IB component is suppressed by 
the $\Delta I = 1/2$ rule, leading to a relative enhancement of the DE 
component. The main contributions to the DE component are the M1 amplitude,
arising from the chiral anomaly, and the E1 component. Both appear at 
$\order{p^4}$ in the chiral expansion. Interference between the M1 and E1 
components vanishes in measurements inclusive with respect to the 
polarization of the radiative photon, but interference between the IB and 
E1 components is in principle observable. The decay is analyzed by looking
at the variables $T^*$, the kinetic energy of the $\pi^\pm$ in the $K^\pm$
CM frame, and $W \equiv (p_K\dot p_\gamma)(p_{\pi^+}\dot p_\gamma)/
(m_{K^+}m_{\pi^+})^2$. For a selected interval in $T^*$, chosen to 
reduce background from other $K^\pm$ decays with $\pi^0$s, the distribution
in $W$ is obtained. This can be written as the sum of a term from IB, a term
from DE, and a term from IB/E1 interference. Fits to the $W$ spectrum 
allow these components to be isolated.

Two $K^+_{\rm stop}$ experiments have recently measured the DE BR for 
$55 < T^* < 90$~MeV: the 2005 preliminary value from BNL E787
\cite{Tsu05:Kaon} is \SN{3.9(5)(^{+3}_{-4})}{-6}; the newly published
value from KEK E470\cite{K470+06:ppg} is \SN{3.8(8)(7)}{-6}.
Both measurements are based on samples of order 10k decays.
No interference term is included in the fit in either case---both 
measurements are improvements of previous analyses in
which no evidence for an interference term was found. 
On the other hand, NA48/2, has a preliminary result based on 
124k in-flight decays from the 2003 data \cite{GL06:Chiral}
obtained with an interference term included in the fit.
The weights of the DE and interference 
terms are 3.35(35)(25)\% and $-2.67(81)(73)\%$. The errors on these two 
values are highly correlated, but the interference term is observed 
with $3\sigma$ significance.
The NA48/2 measurement is for $0<T^*<80$~MeV because of trigger considerations.
For comparison with the $K^+_{\rm stop}$ experiments,
NA48/2 fits with no interference term and extrapolates to
$55 < T^* < 90$~MeV, obtaining a DE fraction of 0.85(5)(2)\%,
which gives a DE BR of about \SN{2.2}{-6}.

For the decay $K_L\to\pi^+\pi^-\gamma$, the IB and E1 amplitudes violate
$CP$, leading to a relative enhancement of the M1 contribution.
KTeV has recently published a new measurement in this 
channel \cite{KTeV+06:ppg} based on 112k events
with $E_\gamma > 20$~MeV, updating their 2001 result
with an increase in statistics
of more than a factor of ten. In fits to the 
$(E_\gamma, \cos\theta_{\pi^+\gamma})$ distribution, the form factor 
describing the M1 amplitude is based on a pole model with VMD photon 
couplings \cite{LV88:ppg}. 
KTeV obtains the best measurement to date of the M1 form-factor parameters, 
as well as the 90\% CL limit
$|g_{\rm E1}|<0.21$. Setting $|g_{\rm E1}|$ to zero, KTeV obtains
the fraction of the radiation spectrum from M1 DE:
${\rm DE/(DE + IB)} = 0.689(21)$ for $E_\gamma>20$~MeV.

KTeV has also recently published a new measurement of the decay
$K_L\to\pi^+\pi^-e^+e^-$ \cite{KTeV+06:ppee}. 
In this decay, the polarization of the virtual 
photon is measured by the plane of the $\gamma^*\to e^+e^-$ conversion,
so that the interference between the IB/E1 and M1 amplitudes
can be observed. In addition, the process contributes
in which the $K_L$ emits a virtual photon and is transformed into a $K_S$, 
which then decays to $\pi^+\pi^-$; the amplitude for this process is 
proportional to \mean{r_K^2}, the charge radius of the neutral kaon.
With $\sim$5200 events, KTeV obtains results for 
\mean{r_K^2} and the M1 form-factor parameters that agree with and improve
upon the previous results from KTeV 
and NA48. 
The M1 form-factor parameter values also agree with
those KTeV obtains for the $K_L\to\pi^+\pi^-\gamma$ channel, while
the new KTeV limit on $|g_{\rm E1}|/|g_{\rm M1}|$ from
$K_L\to\pi^+\pi^-e^+e^-$ is a stronger constraint on the size of the 
E1 amplitude than is obtained from $K_L\to\pi^+\pi^-\gamma$. The asymmetry
$A_\phi$ about zero in the distribution of $\sin\phi\cos\phi$, 
where $\phi$ is the
angle between the $\pi^+\pi^-$ and $e^+e^-$ planes in the decay, 
parameterizes $CP$ violation in the interference between IB and M1 amplitudes.
The new KTeV measurement, $A_\phi = 0.136(14)(15)$,
significantly improves on previous results from KTeV and NA48.

\subsection{$K_S\to\gamma\gamma$}

In ChPT calculations of the amplitude for $K_S\to\gamma\gamma$,
since all particles involved are neutral,
there are no tree-level contributions.
Moreover, at \order{p^4}, only finite chiral-meson loops contribute.
$BR(K_S\to\gamma\gamma)$ is predicted unambiguously at this
level in terms of the couplings $G_8$ and $G_{27}$,
giving \SN{2.1}{-6} \cite{DA+94:Hand2}.
The most precise published measurement of this BR is from
NA48: ${\rm BR}=\SN{2.78(6)(4)}{-6}$ \cite{NA48+03:KSgg}.
This result would suggest the need for a significant \order{p^6} 
correction in the ChPT calculation of the BR.
The NA48 result may soon be confirmed by KLOE \cite{Mar06:Chiral}.
While the number of $K_S\to\gamma\gamma$ events observed by 
KLOE is $\sim$600, as compared to the $\sim7500$ observed by NA48,
KLOE profits from the use of a tagged $K_S$ beam and does not
have to contend with irreducible background from $K_L\to\gamma\gamma$.
The KLOE measurement is in progress; a total error of 5\% is expected,
which is sufficient to confirm the NA48 result.

\section{Determination of \Vusf\ from $K_{\ell3}$ Decays}

A precise test of CKM unitarity can be obtained 
from the first-row constraint
$|V_{ud}|^2 + |V_{us}|^2 + |V_{ub}|^2 = 1$ (with $|V_{ub}|^2$ negligible).
At present, the most precise value for $|V_{us}|$ is obtained from
$K_{\ell3}$ decay rates, via
\begin{equation}
\Gamma(K_{l3(\gamma)})=N_K\:
|V_{us}|^2\:|f_+^{K^0\pi^-}(0)|^2\:I_{K\ell}\:
(1 + 2\Delta_K^{SU(2)} + 2\Delta_{K\ell}^{\rm EM}),
\label{eq:Vus}
\end{equation}
where the subscripts $K$ and $\ell$ indicate dependence on the kaon charge
state ($K^\pm$, $K^0$) and lepton flavor,
and $N_K$ is a well-determined constant. 
The value of the hadronic matrix element 
at zero momentum transfer, $f_+(t=0)$, differs 
from unity because of $SU(2)$- and $SU(3)$-breaking effects; 
conventionally, the
value for $K^0\to\pi^-$ decays is used in \Eq{eq:Vus} and 
$SU(2)$-breaking corrections
are encoded in $\Delta_K^{SU(2)}$. $\Delta_{K\ell}^{\rm EM}$ is the
correction to the form factor for the effects of long-distance electromagnetic
interactions. These theory inputs to \Eq{eq:Vus}, and especially the status
of $f_+^{K^0\pi^-}(0)$, are discussed in the contribution to these 
proceedings by V.~Cirigliano (see also \Ref{Cir06:rad}). 
The inputs from experiment are $\Gamma(K_{l3(\gamma)})$, the 
radiation-inclusive decay rates, or in practice, 
BR and lifetime measurements; and $I_{K\ell}$, the phase-space integrals 
of the form factors, 
which are calculated from measurements of the form-factor slopes 
$\lambda$ as discussed in \Sec{sec:ff}.

In the 2002 PDG evaluation, $|V_{ud}|^2 + |V_{us}|^2 = 0.9965(15)$,
a $2.3\sigma$ hint of CKM unitarity violation.
The 2003 result from BNL E865 \cite{E865+03:Ke3},
$\BR{K^\pm_{e3}} = 5.13(2)(10)\%$, is $2.7\sigma$ higher than the 
2002 PDG average for this BR and seemed to resolve the problem.
With respect to the older measurements of $K_{\ell3}$ decays, 
the E865 result made use of much higher statistics, and in addition
was the first measurement with a well-defined treatment of the 
contribution from radiative decays. All newer measurements from
KTeV, KLOE, ISTRA+, and NA48 share this feature.

\subsection{$K_L$ and $K_S$ branching ratios and lifetimes}
\label{sec:KLfit}

KTeV, NA48, and KLOE have recently published measurements of the BRs
for the dominant $K_L$ decay channels, including the $K_{\ell3}$
decays.

KTeV has measured five ratios of the BRs 
for the principal $K_L$ decays \cite{KTeV+04:BR}.
The six BRs involved account for 99.93\% of the $K_L$ width; 
the ratios are combined to determine the absolute BR values.

NA48 has measured the ratio of $\BR{K_{e3}}$ 
to the sum of the BRs for all decays to two tracks \cite{NA48+04:BR}. 
This is essentially $\BR{K_{e3}}/[1-\BR{3\pi^0}]$. NA48 normalizes using 
the average of the KTeV and NA31 measurements of $\BR{3\pi^0}$;
they also have a preliminary measurement of $\BR{3\pi^0}$
normalized to $K_S\to\pi^0\pi^0$ decays that confirms this average
\cite{Lit04:ICHEP}.

KLOE has measured
absolute BRs for the four dominant $K_L$ modes
using $\phi\to K_S K_L$ 
events in which a $K_S\to\pi^+\pi^-$ decay is used to tag the $K_L$ decay, 
providing normalization \cite{KLOE+06:BR}. 
The dominant contribution to the uncertainties on
the absolute BRs comes from the uncertainty on $\tau_L$, the $K_L$ lifetime,
which is needed to calculate the overall geometrical efficiency.
Expressing the BRs as functions of $\tau_L$ and imposing the
constraint $\sum {\rm BR} = 1$ (with the 2004 PDG values used for the 
smaller $K_L$ BRs), final
values are obtained for the four BRs and for $\tau_L$, with
greatly reduced uncertainties. 
KLOE has also measured $\tau_L$ directly, using $10^7$ $K_L\to3\pi^0$ events
\cite{KLOE+05:tauL}, for which the reconstruction efficiency
is high and uniform
inside the fiducial volume ($0.37 \lambda_L$). 
The result, $\tau_L = 50.92(17)(25)$~ns, 
is consistent with the value obtained from the sum of the $K_L$ BRs.

The results from the experiments are summarized in \Tab{tab:KLBR}. 
Because of their interdependence for the purposes of normalization, 
they are best incorporated into a new evaluation of \Vusf\ via
a global fit akin to that performed by the PDG. 
\begin{table}
\tbl{Recent $K_L$ BR and lifetime measurements used in fit}
{\begin{tabular}{@{}lll@{}}
\toprule
KTeV$^{\text a}$ & 
     \BR{K_{\mu3}/K_{e3}} = 0.6640(26) &
     \BR{3\pi^0/K_{e3}} = 0.4782(55) \\
&    \BR{\pi^+\pi^-\pi^0/K_{e3}} = 0.3078(18) &
     \BR{\pi^+\pi^-/K_{e3}} = \SN{4.856(29)}{-3} \\
&    \BR{\pi^0\pi^0/3\pi^0} = \SN{4.446(25)}{-3} \\ \colrule
KLOE$^{\text{a,b}}$ & 
     \BR{K_{e3}} = 0.4049(21) &
     \BR{K_{\mu3}} = 0.2726(16) \\
&    \BR{3\pi^0} = 0.2018(23) &
     \BR{\pi^+\pi^-\pi^0} = 0.1276(15) \\
&    $\tau_L$ = 50.92(30)~ns \\ \colrule
NA48 & 
     \BR{K_{e3}/{\mbox{2 track}}} = 0.4978(35) & 
     \BR{3\pi^0} = 0.1966(34)$^{\text c}$ \\
\botrule
\end{tabular}}
\begin{tabnote}
$^{\text a}$ In the fit, errors on these BRs are parameterized by the complete 
covariance matrix.
$^{\text b}$ In the fit, these BR values are expressed as 
functions of $\tau_L$ as described in \Ref{KLOE+06:BR}.
$^{\text c}$ Preliminary.\\
\end{tabnote}
\label{tab:KLBR}
\end{table}
The fit performed here uses the data in \Tab{tab:KLBR} in 
addition to four other measurements used in the 2006 PDG fit.
The free parameters are the seven largest $K_L$ BRs and $\tau_L$;
the BRs in the fit are constrained to to sum to unity.
The principal difference between the fit performed here and the 2006 PDG fit
is that here, the intermediate KTeV and KLOE values 
(\ie, before applying constraints) are the inputs, and the complete 
error matrix is used to handle the correlations between the measurements 
from each experiment.
(In the 2006 PDG fit, the final KTeV and KLOE BR results were used
and one measurement involving $\BR{3\pi^0}$ was removed in each case.)
Scale factors for the errors are calculated and used as 
per the PDG prescription.
The fit has $\chi^2/{\rm ndf} = 13.2/9$ (15.4\%) and gives
$\BR{K_{e3}} = 0.4047(11)$ $(S=1.4)$, 
$\BR{K_{\mu3}} = 0.2698(9)$ $(S=1.3)$, and
$\tau_L = 51.11(21)$~ns $(S=1.1)$. 
These values are quite similar to those from the 2006 PDG fit.

KLOE also has recently published \cite{KLOE+06:KSe3} 
a measurement of \BR{K_S\to\pi e\nu} that is precise enough 
to contribute meaningfully to the evaluation of \Vusf. 
For this measurement, $K_S$ decays are tagged by the observation
of a $K_L$ interaction in the KLOE calorimeter.
The quantity directly measured is \BR{\pi e\nu/\pi^+\pi^-}. Together 
with the recently published KLOE value \BR{\pi^+\pi^-/\pi^0\pi^0} = 2.2459(54),
the constraint that the $K_S$ BRs must sum to unity, and the assumption of 
universal lepton couplings, this completely determines the $K_S$ BRs for 
$\pi^+\pi^-$, $\pi^0\pi^0$, $K_{e3}$, and $K_{\mu3}$ decays 
\cite{KLOE+06:Kspp}. In particular, $\BR{K_S\to\pi e\nu} = \SN{7.046(91)}{-4}$.
The KLOE measurement is performed separately for
each lepton charge state, yielding the first result for the semileptonic
charge asymmetry from $K_S$ decays, $A_S = \SN{15(96)(29)}{-4}$. 
This value has been used in tests of $CPT$ symmetry and the 
$\Delta S = \Delta Q$ rule \cite{KLOE+06:BSR}. 

\subsection{$K^\pm$ branching ratios and lifetime}

NA48/2, ISTRA+, and KLOE all have preliminary measurements of $K^\pm$ BRs.
These new measurements have a
significant impact on the evaluation of \Vusf, as demonstrated by 
the updated fit performed here. 

NA48/2 measures \BR{K_{e3}/\pi\pi^0} and uses the 2004 PDG value
for \BR{\pi\pi^0} to quote \BR{K_{e3}} = 5.14(2)(6)\% \cite{Lit04:ICHEP}.
The fit performed here makes use of the value 
\BR{K_{e3}/\pi\pi^0} = 0.2433(25), as well as of the NA48/2 result
\BR{K_{\mu3}/K_{e3}} = 0.6749(35)(24).

ISTRA+ also measures \BR{K_{e3}/\pi\pi^0}; they use the 2006 PDG value
for \BR{\pi\pi^0} to quote \BR{K_{e3}} = 5.170(11)(57)\% \cite{Duk06:ICHEP}.
The fit performed here makes use of the value 
\BR{K_{e3}/\pi\pi^0} = 0.2471(23).

KLOE measurements the absolute $K_{e3}$ and $K_{\mu3}$ BRs
\cite{Sci06:Lisbon}.
In $\phi\to K^+ K^-$ decays, $K^+$ decays into $\mu\nu$ or $\pi\pi^0$
are used to tag a $K^-$ beam, and vice versa. KLOE performs four 
separate measurements for each $K_{\ell3}$ BR, corresponding to the
different combinations of kaon charge and tagging decay.
The final averages are \BR{K_{e3}} = 5.047(46)(80)\% and
\BR{K_{\mu3}} = 3.310(40)(70)\%.
The fit performed here takes into
account the dependence of these BRs on the $K^\pm$ lifetime.

The world average value for $\tau_\pm$ is nominally 
quite precise; the 2006 PDG quotes $\tau_\pm = 12.385(25)$~ns.
However, the error is scaled by 2.1; the confidence level for the 
average is 0.2\%. It is important to confirm the value of $\tau_\pm$.
KLOE has a preliminary measurement based on $K^\pm$ decays tagged
by $K^\mp \to \mu\nu$ and observed in 
a fiducial volume of $\sim$$1\lambda_\pm$ \cite{Pal06:Moriond}. The result,
$\tau_\pm = 12.336(44)(65)$~ns, agrees with the PDG average, although
at present the KLOE uncertainty is significantly larger.

\begin{figure}
\begin{center}
\psfig{file=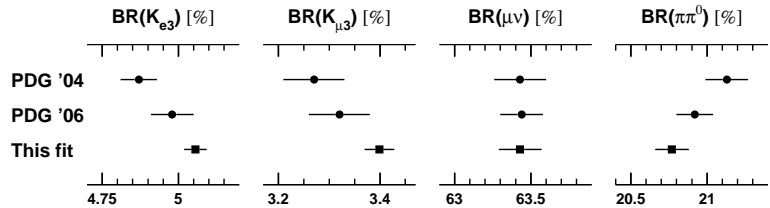,width=4in}
\end{center}
\caption{Evolution of average values for main $K^\pm$ BRs.}
\label{fig:kpmavg}
\end{figure}

The fit performed here makes use of all preliminary results cited above, 
plus the data used in the 2006 PDG fit, for a total of 30 measurements.
The free parameters are the six main $K^\pm$ BRs and $\tau_\pm$; 
the BRs are constrained to sum to unity.
The fit gives $\chi^2/{\rm ndf} = 38/24$ (3.6\%).
The poor fit quality principally derives from the 
scatter in the five older measurements of $\tau_\pm$; when these are 
replaced with their PDG average with scaled error, $\tau_\pm = 12.385(25)$~ns,
the fit gives $\chi^2/{\rm ndf} = 20.5/20$ (42\%), with no significant
changes in the results.
The results are 
$\BR{K_{e3}} = 5.056(37)\%$ $(S=1.3)$, 
$\BR{K_{\mu3}} = 3.399(29)\%$ $(S=1.2)$, and
$\tau_\pm = 12.384(21)$~ns $(S=1.8)$. 
The significant evolution of the average values of the BRs for 
$K^\pm_{\ell3}$ decays and for 
the important normalization channels is evident in \Fig{fig:kpmavg}. 

\subsection{$K_{\ell3}$ form-factor slopes}
\label{sec:ff}
Only the vector part of the weak current contributes to the hadronic
matrix element for $K_{\ell3}$ decays:
\begin{displaymath}
\bra{\pi}J_\alpha\ket{K} = f_+(t)(p_K+p_\pi)_\alpha + f_-(t)(p_K-p_\pi)_\alpha,
\end{displaymath}
where $t=(p_K-p_\pi)^2$.
When the squared matrix element is evaluated, a factor of $m_\ell^2/m_K^2$
multiplies all terms containing $f_-(t)$. This form factor can be 
neglected for $K_{e3}$ decays. For the description of $K_{\mu3}$ decays,
it is customary to use $f_+(t)$ and 
the scalar form factor $f_0(t) \equiv f_+(t) + [t/(m_K^2-m_\pi^2)]\,f_-(t)$. 
The form factors are written as 
$f_{+,\,0}(t)=f_+(0)\tilde f_{+,\,0}(t)$, with $\tilde f_{+,\,0}(0)=1$,
and often expanded in powers of $t$ as
\begin{equation}
\begin{array}{lcll}
\tilde f(t) & = & 1 + \lambda\frac{t}{m^2_{\pi^+}} &
\hspace{1cm}\mbox{(linear),}\\
\tilde f(t) & = & 1 + \lambda'\frac{t}{m^2_{\pi^+}} +
\frac{1}{2}\lambda''\left(\frac{t}{m^2_{\pi^+}}\right)^2 &
\hspace{1cm}\mbox{(quadratic).}\\
\end{array}
\label{eq:FF}
\end{equation}
The slopes $\lambda$ are obtained from fits to the measured $t$ distributions,
but sensitivity to the quadratic terms is poor, in large part because the 
kinematic density of the matrix element drops to zero at large $t$, 
where the form factor itself is maximal.  
The vector form factor $f_+$ is dominated by the vector $K\pi$ resonances,
\eg, $K^*(892)$; this fact suggests the pole parameterization,
$\tilde f_+(t)=M_V^2/(M_V^2-t)$.
This one-parameter form generally fits
experimental data better than does the linear parameterization;
its expansion gives $\lambda'=(m_{\pi^+}/M_V)^2$;
$\lambda''=2\lambda'^2$. A dispersive representation for $f_0(t)$
featuring a single experimental parameter has also 
been proposed\cite{Ste06:Chiral}.  

KTeV, KLOE, and NA48 have measured the form-factor slopes 
in $K_L$ decays; ISTRA+ has measured the slopes in $K^-$ decays.
KTeV and ISTRA+ have reported fit results for 
$\lambda_+'$, $\lambda_+''$, and $\lambda_0$ 
for both $K_{e3}$ and $K_{\mu3}$ decays;
KLOE and NA48 have values for $\lambda_+'$ and $\lambda_+''$ from
$K_{e3}$ decays. These data are collected in \Tab{tab:FF}.
The experiments use different conventions in reporting the 
form-factor slopes; the data in the table have been adjusted
for use with \Eq{eq:FF}.
Most experiments also quote various other combinations of linear,
quadratic, and pole fit results. NA48 has preliminary results
for $\lambda_+$ and $\lambda_0$ (linear fit) from $K_{\mu3}$ decays
\cite{Win05:Lisbon}.
\begin{table}
\tbl{Measurements of $K_{\ell3}$ form-factor slopes}
{\begin{tabular}{@{}lccc@{}}
\toprule
Experiment & \SN{\lambda_+'}{3} & \SN{\lambda_+''}{3} & \SN{\lambda_0}{3} \\
\colrule
KTeV $K_L$ $e3$-$\mu3$ avg. (\Ref{KTeV+04:FF}) &
     $20.6\pm1.8$ & $3.2\pm0.7$ & $13.7\pm1.3$ \\
KLOE $K_L$ $e3$ (\Ref{KLOE+06:FF}) & 
     $25.5\pm1.8$ & $1.4\pm0.8$ \\
NA48 $K_L$ $e3$ (\Ref{NA48+04:FF}) &
     $28.0\pm2.4$ & $0.4\pm0.9$ \\
ISTRA+ $K^-$ $e3$ (\Ref{ISTRA+04:e3FF}) & 
     $24.9\pm1.7$ & $1.9\pm0.9$ \\
ISTRA+ $K^-$ $\mu3$ (\Ref{ISTRA+04:m3FF})$^{\text a}$ & 
     $23.0\pm6.4$ & $2.3\pm2.3$ & $17.1\pm2.2$ \\
\botrule
\end{tabular}}
\begin{tabnote}
$^{\text a}$ No systematic uncertainties are quoted for this fit. \\
\end{tabnote}
\label{tab:FF}
\end{table}

To obtain reference values of the form-factor slopes for the
phase-space integrals, the data in \Tab{tab:FF} have been averaged.   
Correlation coefficients are available for the KTeV and KLOE data;
for NA48 and ISTRA+, their values have been 
inferred, in part by assuming the correlations to be intrinsic to the
measurement and largely independent of the experimental details.
In principle, this fact could be exploited to fix all
correlations {\it a priori} \cite{Fra06:FFnote}.
The results are
$\lambda_+'=\SN{24.72(84)}{-3}$,
$\lambda_+''=\SN{1.67(36)}{-3}$, and
$\lambda_0=\SN{15.72(97)}{-3}$, with
$\rho(\lambda_+',\lambda_+'')=-0.94$,
$\rho(\lambda_+',\lambda_0)=+0.30$, and
$\rho(\lambda_+'',\lambda_0)=-0.40$.
The fit gives $\chi^2/{\rm ndf}=11.6/9$ (23.9\%).

KTeV, KLOE, and NA48 all quote values for $M_V$ for $K_{e3}$ decays.
The average value is $M_V=875.3\pm5.4$~MeV with $\chi^2/{\rm ndf}=1.83/2$
(40\%).
Using this or the avergage of the quadratic fit results stated above to 
calculate the phase-space integral for the $K_{L\,e3}$ form factor
makes a 0.02\% difference in the result. In the calculation of \Vusf,
no additional error is assigned to account for differences obtained 
with quadratic and pole parameterizations for $\tilde f_+(t)$. 

\subsection{Discussion}
\begin{figure}
\center
\begin{minipage}{0.92\textwidth}
\psfig{file=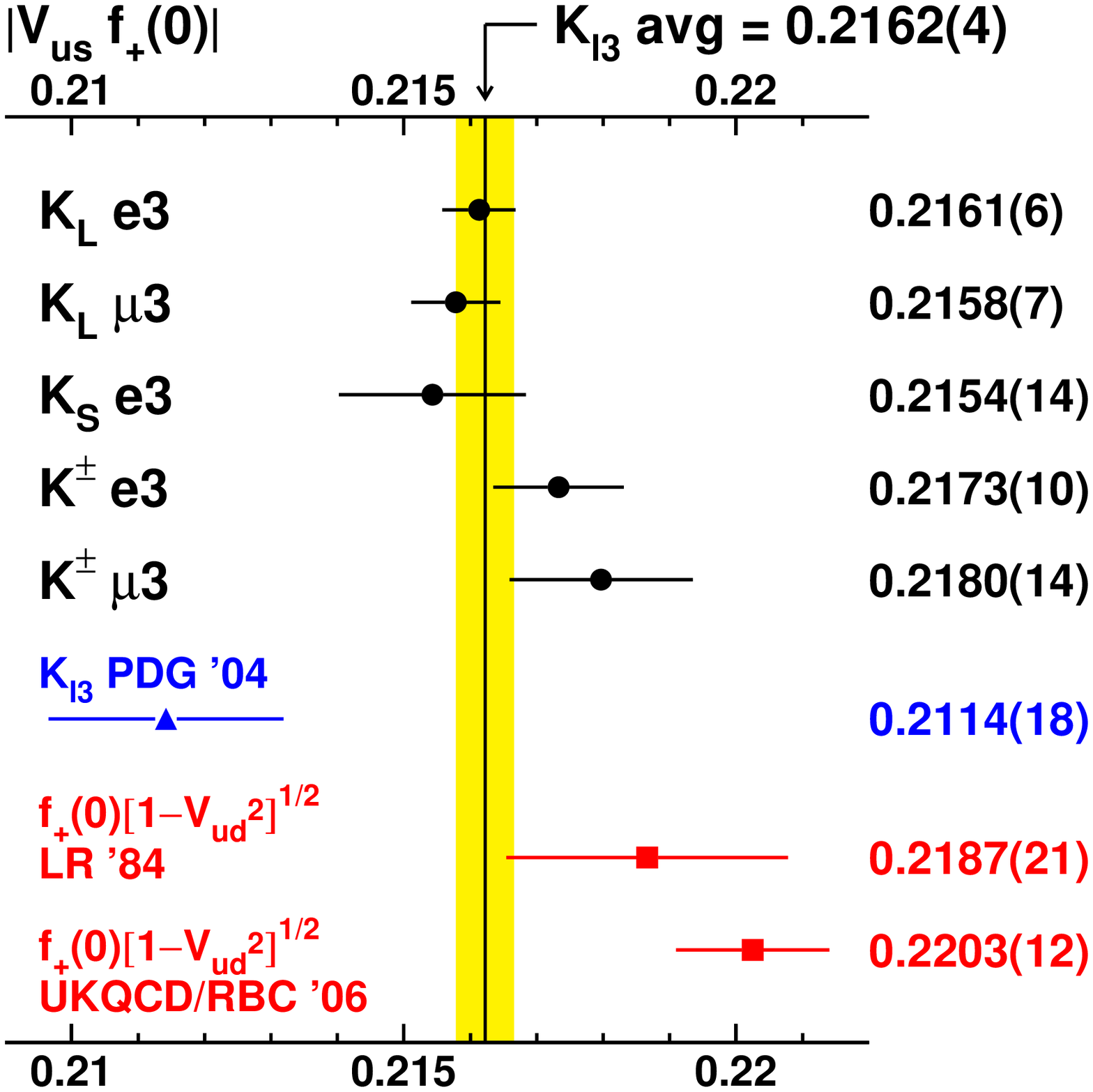,height=1.8in}
\hspace{\fill}
\psfig{file=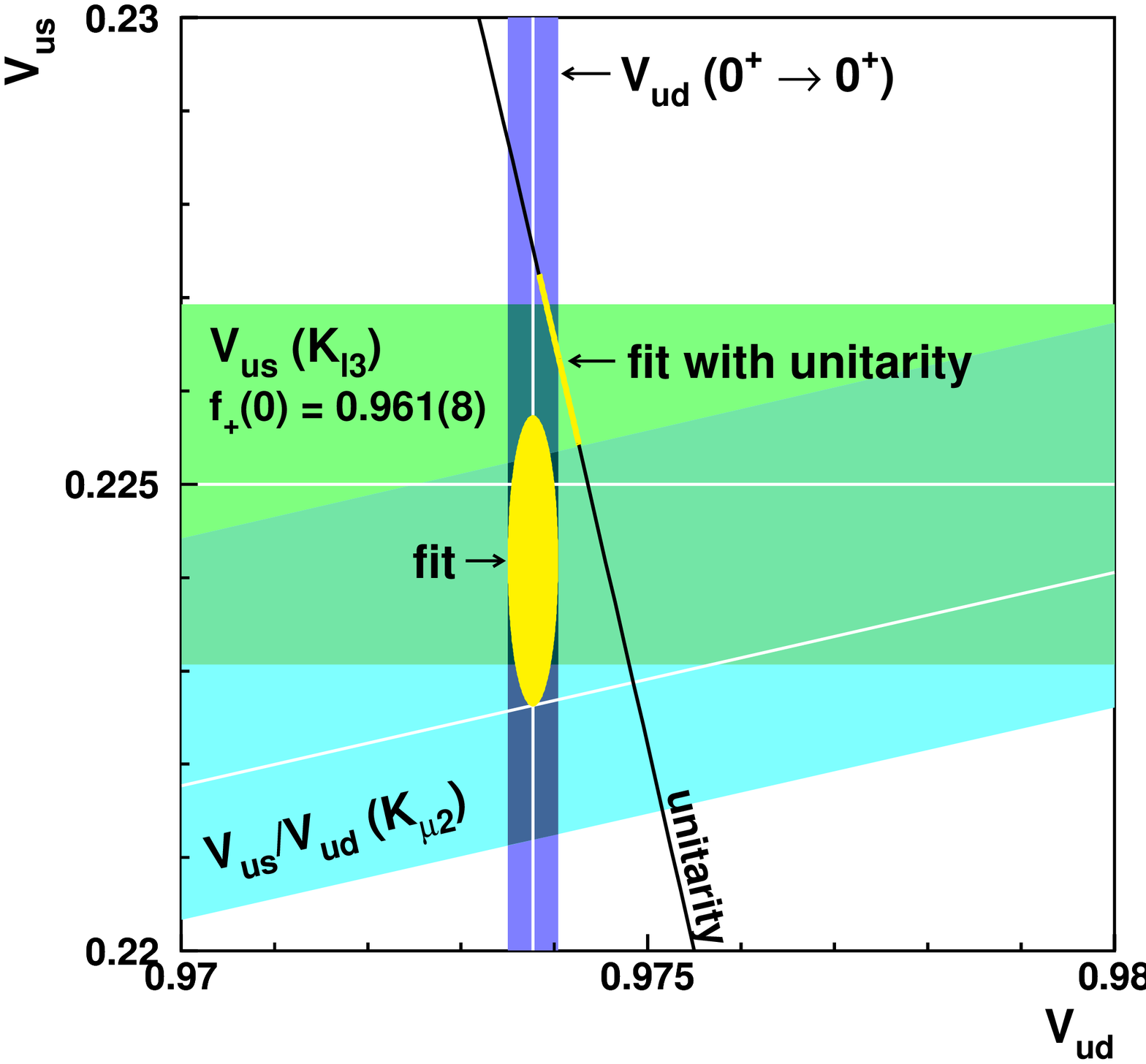,height=1.8in}
\end{minipage}
\caption{Left: Determinations of \Vusf\ for five $K_{\ell3}$ decay channels,
with average and comparison values. Right: Results of a fit to values 
for $|V_{ud}|$, $|V_{us}|$, and $|V_{us}|/|V_{ud}|$.}
\label{fig:vus}
\end{figure}

Using the results of the fits discussed above for the BRs, lifetimes, and 
form-factor slopes, \Vusf\ has been evaluated for each of the five decay 
modes using \Eq{eq:Vus}.
The results are summarized in the left panel of \Fig{fig:vus}.
The most precise determination is from $K_{L\,e3}$ decays:
$\Vusf = 0.2161(6)$. Indicatively, for $K_L$ decays, the precison is
limited by the uncertainties on the decay widths, and in particular, by the
uncertainty 
on $\tau_L$. For the $K_{L\,\mu3}$ mode, the uncertainty on
$\Delta^{\rm EM}$ is also a significant issue. For both $K^\pm$ decays, 
the uncertainties on the BR measurements are the limiting factor;
for $K^\pm_{\mu3}$, the uncertainty on $\Delta^{\rm EM}$ is nearly 
as important. Although not a limiting factor at the moment, for $K^\pm$ 
decays, the
present $\sim$10\% uncertainty on $\Delta_{K^\pm}^{SU(2)}$ would ultimately
prohibit obtaining the same level of precision as for $K_L$ decays.
The uncertainty on the phase-space integral is not 
currently a limiting factor for any mode.

A fit to all five values for \Vusf\ taking 
all correlations into account has $\chi^2/{\rm ndf} = 2.76/4$ (60\%)
and gives the average value $\Vusf=0.2162(4)$. 
When \Vusf\ is evaluated 
separately for $K^0$ and $K^\pm$ decays (making use of separate averages
for the form-factor slopes for each case), the result for $K^\pm$ decays
is $1\sigma$ higher than that for $K^0$ decays. At present, the results
obtained from all modes are consistent, though it is worth noting that
all of the new $\BR{K^\pm_{\ell3}}$ results are still preliminary. 

To test CKM unitarity, a value for $f_+(0)$ is needed. 
Conventionally, the original estimate of Leutwyler and Roos \cite{LR84:f0},
$f_+(0) = 0.961(8)$, is used; this gives $|V_{us}|=0.2250(19)$.
Using the most recent evaluation of $|V_{ud}|$ from $0^+\to0^+$ 
nuclear beta decays \cite{MS06:Vud}, $|V_{ud}| = 0.97377(27)$,
one has $|V_{ud}|^2 + |V_{us}|^2 - 1 = -0.0012(10)$, a result
perfectly compatible with unitarity. 
As is evident from \Fig{fig:vus},  
this represents a significant 
evolution of the experimental picture since the 2004 PDG evaluation.
However, lattice evaluations of $f_+(0)$ are rapidly improving
in precision.
For example, the UKQCD/RBC Collaboration has announced a preliminary 
result \cite{A+06:f0lat} from a lattice calculation with $2+1$ flavors of 
dynamical domain-wall quarks: $f_+(0) = 0.9680(16)$. This value implies
$|V_{us}|=0.2234(6)$ and $|V_{ud}|^2 + |V_{us}|^2 - 1 = -0.0019(6)$,
a $3.2\sigma$ discrepancy with CKM unitarity.

Marciano \cite{Mar04:fKpi} has observed that 
$\Gamma(K_{\mu2})/\Gamma(\pi_{\mu2})$ can be precisely related to 
the product $(|V_{us}|/|V_{ud}|)^2(f_K/f_\pi)^2$. The recent
measurement $\BR{K^+\to\mu^+\nu}=0.6366(9)(15)$ from KLOE \cite{KLOE+06:Km2},
together with the preliminary lattice result 
$f_K/f_\pi = 1.208(2)(^{+7}_{-14})$ from the MILC Collaboration 
\cite{Ber06:Chiral}, gives $|V_{us}|/|V_{ud}| = 0.2286(^{+27}_{-15})$.
This ratio can be used in a fit together with the values of $|V_{ud}|$
from \Ref{MS06:Vud} and $|V_{us}|$ from $K_{\ell3}$ decays as above.
Using the value for $|V_{us}|$ obtained with $f_+(0)=0.961(8)$, 
the fit gives $|V_{ud}| = 0.97377(27)$ and $|V_{us}| = 0.2242(16)$, 
with $\chi^2/{\rm ndf} = 0.52/1$ (47\%). 
The unitarity constraint can also be included, in which 
case the fit gives $\chi^2/{\rm ndf} = 3.43/2$ (18\%).
Both results are illustrated in \Fig{fig:vus}, right.
If instead the newer lattice 
result for $f_+(0)$ is used to obtain $|V_{us}|$,
the fit gives $|V_{ud}| = 0.97377(27)$ and $|V_{us}| = 0.2233(6)$, 
with $\chi^2/{\rm ndf} = 0.074/1$ (79\%). When the unitarity constraint
is imposed, the fit gives $\chi^2/{\rm ndf} = 10.5/2$, corresponding
to a probability of 0.53\%.
These results reinforce the conclusion that 
the result of the first-row test of CKM unitarity depends 
mainly on the value and uncertainty assumed for 
$f_+(0)$. Confirmation of the new lattice result is a critical step 
towards a clearer understanding of the situation.

\section*{Acknowledgments}
I would like to congratulate the members of the E470, E787, E865, KTeV, 
ISTRA+, and NA48 collaborations, as well as my fellow KLOE collaborators,
for their hard work. I apologize for omitting many results
for reasons of space. 
Special thanks go to 
M.~An\-to\-nel\-li, 
V.~Ci\-ri\-glia\-no, and 
P.~Fran\-zi\-ni for many useful discussions; to
B.~Scia\-scia for help with the fits; and to
G.~Is\-i\-do\-ri and T.~Spa\-da\-ro 
for comments on the manuscript.

\bibliographystyle{ws-procs9x6}
\bibliography{moulson}

\end{document}